\documentclass[epj,referee]{svjour}
\usepackage{graphics}
\usepackage{graphicx}
\usepackage{textcomp}
\begin{document}
\title{On fundamental mechanisms in dye sensitized solar cells through the behaviour of different mesoporous titanium dioxide films}
\subtitle{Download here   https://hal.archives-ouvertes.fr/hal-01137249v2}

\author{L\'idice Vaillant \inst{1} \and Elena Vigil\inst{1,2} \and Fresnel Forcade \inst{1} \and Thierry Thami \inst{3} \and Hania Adnani \inst{4} \and Christelle Yacou \inst{3} \and Andr\'e Ayral \inst{3} \and Pierre Saint-Gr\'egoire \inst{5}
}                     
\offprints{}          
%
\institute{ Enermat Division, Institute of Materials Science and Technology (IMRE) - Physics Faculty, University of Havana, Zapata esq G, s/n, 10400 Cuba \and Physics Faculty, University of Havana, San L\'azaro y L, 10400 Cuba \and Institut Europ\'een des Membranes, UMR 5635 CNRS, ENSCM, Universit\'e de Montpellier, CC047 – Place Eug\`ene Bataillon, 34095 Montpellier C\'edex 5, France \and University Ferhat Abbhas, S\'etif 1, Algeria \and MIPA Laboratory, University of N\^imes, Dept Sciences and Arts, CS 13019, 30021 N\^imes cedex 01, and University of Toulon, France 
\\Corresponding author : pstgregoire@gmail.com}
\date{Received: date / Revised version: date}
%
\abstract{
Understanding mechanisms in DSSCs is fundamental for their improvement; this includes the nanocrystalline semiconducting layer behaviour. Different mesoporous $TiO_2$ layers are fabricated and analyzed for possible use in DSSC solar cells. The preparations included the addition of P123 triblock copolymer as structuring agent to the synthesized anatase sol. This preparation was also mixed with P25 $TiO_2$ nanoparticles in one case and polystyrene latex in another. Mesoporous mixed $TiO_2-SiO_2$ thin layers were also analyzed. The diverse morphology and features are studied by microscopic techniques and by means of spectral quantum efficiency of a photoelectrochemical cell (PEC) that uses as photoelectrode the unsensitized porous $TiO_2$ layer. Contact angle measurements are also performed. We have found that a very high specific area due to very small nanocrystals and small pores can hinder electrolyte penetration in the pores formed by $TiO_2$ nanograins, affecting photoelectrodes efficiency. 
} 
\maketitle
\section{Introduction}
\label{intro}
Study and development of dye sensitized solar cells (DSSCs) is an up-to-date field of research.  The main prospective of these devices is based on the possibility of achieving very inexpensive and efficient solar conversion. Fabrication of DSSCs is simpler than silicon or any thin film based solar cells, where technological steps require high temperatures, control of impurity concentration and high vacuum deposition process. Moreover, the conception of a three-dimensional (3D) interface between active materials and the separation of roles of the absorber and the transporting materials are distinctive differences with traditional solar cells \cite{Green91}. 
The fundamental structure of the cell remains essentially unchanged since its first appearance in scientific publications \cite{Regan91}. The cell consists of three major important parts, the front electrode of nanostructured $TiO_2$, the dye for light absorption and the electrolyte. After absorption of sunlight, the dye injects electrons to the $TiO_2$ semiconductor, where they are transported by diffusion \cite{Regan91}. The electrolyte regenerates the oxidized dye and transports holes to the counter-electrode.
According to the role of nanostructured $TiO_2$ in DSSC, one important issue is obtaining $TiO_2$ films with adequate porosity and crystallinity at low temperature, preferably using simple technology. $TiO_2$ nanocrystalline and porous character determines a different working principle in DSSC when compared to traditional solar cells based on flat junctions. The nano-sized three-dimensional interface structure for the electron conducting medium (most frequently porous nanocrystalline $TiO_2$) has different implications: first, light absorption by the sensitizer or dye is enormously enhanced since the real internal area of the porous $TiO_2$ with the attached sensitizer is orders of magnitude larger than the cell area. Second, it allows the electrolyte or other hole conducting medium to penetrate and surround the $TiO_2$ nanocrystals. Therefore, a three-dimensional (3D) heterostructure is formed. Third, it allows charge carriers to be transported in a medium different from that in which photon absorption creates them. Fourth, only diffusion currents are present because due to nanocrystal size there are no macroelectric fields that could generate drift currents. Fifth, it is a majority carrier device. There are no excess minority carriers (holes) in the $TiO_2$ where electrons are transported. Electron-hole recombination does not play an important role in decreasing efficiency as in traditional solar cells. It is the main reason why no highly pure and perfect crystals are required. This causes that DSSCs are quite cheaper than traditional solar cells. Sixth, light suffers multiple scattering inside the 3-D heterostructure. All of these six phenomena determine the DSSC efficiency and all of them depend on the nanostructure of the employed $TiO_2$, e.g., nanocrystal size, porosity, pore size, nanocrystal shape \cite{Vivero12}. These, in turn, depend on the technology used in obtaining the mesoporous $TiO_2$ electron-conducting medium. Recently, several other techniques have been used for the $TiO_2$ film synthesis, including electrochemical anodization for obtaining arrays of nanorods \cite{Djam13}, and miscellaneous techniques for multiple layers \cite{Zhang13}. However up to now, quantitative or semiquantitative analyses of the phenomena in the layers related to porosity, connectivity, and pore size, are rather scarce in literature.
Because of the mentioned reasons, we analyze for possible use in DSSCs, some previously reported techniques for obtaining mesoporous $TiO_2$ for photocatalysis. A simple sol-gel route for the synthesis at low temperature of mesoporous and nanocrystalline anatase thin films from titanium isopropoxide was reported by Bosc et al \cite{Bosc03}. The optimization of the synthesis parameters enabled the room temperature preparation of clear sols consisting of dispersed anatase nanocrystallites \cite{Bosc03}. By the addition of a templating agent to the starting sols, the porosity and specific surface area were controlled, and layers exhibiting ordered mesoporosity were prepared \cite{Bosc03}. Also a method was developed for the preparation at low temperature of mesoporous mixed $TiO_2$-$SiO_2$ thin layers \cite{Bosc06}. The effect of addition of silica on the crystalline structure, on the mesostructure, on the porosity and on the photocatalytic activity was investigated. Nanocomposite solids consisting of amorphous silica and anatase nanocrystals were produced \cite{Bosc06}. For $80\%$ molar titania content, no significant decrease of photocatalytic activity was found compared to pure anatase \cite{Bosc06}. Ayral and coworkers \cite{Bosc06b} also prepared stable complex organic-inorganic hybrid suspensions by mixing a polystyrene latex aqueous suspension with the titania hydrosol containing the nonionic triblock copolymer. These suspensions were deposited as thin films. Solvent evaporation induces the formation of spherical micelles by self-assembly of the amphiphilic molecules during the drying of the films. Two types of isolated spherical macropores (few ten nanometers) and mesopores (4-5 nm) were reported to be generated inside the layers by the thermal removal of the polystyrene particles and of the micelles, respectively \cite{Bosc06b}. The remaining inorganic network exhibits an additional interconnected microporosity with a mean pore size of 1.5 nm, resulting from the aggregation of the anatase nanoparticles \cite{Bosc06b}. 
In the literature, DSSCs are studied as a complete device, i.e., sealed and with sensitization. Generally, analysis of the $TiO_2$ used for making the DSSC is not performed independently - as we do and consider relevant to do \cite{Zumeta02}. In order to characterize the $TiO_2$ porous samples obtained by different synthesis routes we propose to use spectral photocurrent response and spectral quantum efficiency of a photoelectrochemical cell (PEC) that uses as photoelectrode the unsensitized porous $TiO_2$ layer. Advantages of these techniques to characterize the porous $TiO_2$ film for DSSC are discussed. The mentioned techniques are complemented with contact angle measurements. Optical microscopy and scanning electron microscopy (SEM) are also used  to better understand the relation between texture at different scales and physical properties.
\section{Experimental}
\label{sec:1}
\subsection{Technology used for each of the different samples studied}
We have used methods reported by Bosc et al in \cite{Bosc03}, \cite{Bosc06} and \cite{Bosc06b} to obtain mesoporous $TiO_2$ in order to study possible application in DSSC. A fourth suspension type was also used for deposition of thin films, the sol-gel reported in \cite{Bosc03} mixed with nanocrystalline $TiO_2$ P-25 powder supplied by Evonik. 
\subsubsection{Samples labeled A}
Ti(IV) isopropoxide was used as $TiO_2$ precursor. The titanium isopropoxide was hydrolyzed under vigorous stirring by the addition of an aqueous solution of hydrochloric acid. Solutions were prepared in the following conditions: the titanium concentration in the sol, $[Ti] = 0,023 \ mol L^{-1}$; the $HCl/Ti$ ratio, $a = 1$; the $H_2O/Ti$ hydrolysis ratio, $h = 20$. Stirring at constant temperature $T = 30 \ ^oC$\ occurred during $t = 3$ $h$ $30 \ min$ to obtain the acidic anatase hydrosol.
After this time, to generate ordered mesoporosity by the templating effect, the structuring agent (the triblock copolymer poly(ethylene oxide)-poly(propylene oxide)-poly-(ethylene oxide), EO20PO70EO20, labeled P123) was added under stirring at $T = 30\ ^oC$; which was continued during $30 \ min$. The added amount of this large amphiphilic molecule was defined by the volume fraction of the copolymer in the dried layer, assuming that the inorganic phase is pure anatase. The added amount of P123 corresponds to a volume fraction of $70\%$. Films were deposited on conducting $15\  \Omega/ square$ FTO glass TCO22-15 of Solaronix by “doctor-blading”.
\subsubsection{Samples labeled D}
This preparation intended, by adding P-25, to change some characteristics of the $TiO_2$ film, like porosity and to introduce scattering centers to increase the light path in the active layer. An aqueous suspension of  P-25 was prepared following the procedure described in \cite{Bosc03}. After obtaining the sol-gel described for samples “A” (titania hydrosol with the triblock copolymer), it was mixed with the $TiO_2$ P-25 suspension. Mixed volumes were calculated to obtain in the films $20\%$ of $TiO_2$ nanocrystals coming from the titania nanopowder P-25. Films were deposited on conducting $15\  \Omega/ square$ FTO glass TCO22-15 of Solaronix by “doctor-blading”.
\subsubsection{Samples labeled S}
For the preparation at low temperature of mesoporous mixed $TiO_2$-$SiO_2$ thin layers \cite{Bosc06}, tetraethoxysilane (TEOS) diluted in ethanol was used as silica precursor. The procedure described in II.1.1 was followed to obtain the acidic anatase hydrosol. After stirring for 3.5 h, the obtained anatase hydrosol was mixed with the TEOS ethanolic solution under vigorous stirring. This favors a fast hydrolysis of the silicon alkoxide. Concentrations were calculated in order to obtain $80\%$ molar $TiO_2$ content and $20\%$ $SiO_2$ molar content. The structuring agent P123 is finally added as previously explained. Films were deposited on conducting $15\  \Omega/ square$ FTO glass TCO22-15 of Solaronix by “doctor- blading”.
\subsubsection{Samples labeled L}
The last preparation sought the presence of macropores ($\sim$ 100 nm) and mesopores ($\sim$ 4-5 nm) in the porous $TiO_2$ films. After obtaining the sol-gel described for samples “A” (titania hydrosol with the triblock copolymer), it was mixed with a polystyrene latex aqueous suspension ($20\%$ molar concentration). Polystyrene particles had a diameter $<\phi>\ \sim 90\  nm$ and they create macropores when thermally removed. After mixing, stirring proceeded for $30 \ min$ at $30\ ^oC$. Films were deposited by “dip-coating” (withdrawal rate = 5 inches/min) on conducting $15\  \Omega/ square$ FTO glass TCO22-15 of Solaronix.
All film types were dried at room temperature and under controlled humidity (relative humidity $RH = 70\%$) for over 24 hours. After this, they were pre-treated thermally in an air atmosphere: temperature was increased at $0.5\ ^oC/min$ and kept constant during $6 h$ at $100\ ^oC$ and $6 h$ at $150\ ^oC$. After cooling to room temperature ($15\ ^oC/min$), the films were calcinated. Temperature increased $2\ ^oC/min$ up to $150\ ^oC$ and at $0.5\ ^oC/min$ up to $250\ ^oC$. Temperature was kept constant for one hour at $250\ ^oC$ and then the temperature was increased at $0.5\ ^oC/min$ up to $350\ ^oC$. After a $2 h$ heat-treatment, temperature was increased to $450\ ^oC$. Samples were heat-treated for one hour at this temperature and then cooled back to room temperature at $20\ ^oC/min$. 
\subsection{Characterization techniques}
Sample surface preliminary observations were performed with a polarized light metallographic microscope NJF-1 with a 1 $\mu m$ resolution. For scanning electron microscopy (SEM) a Hitachi S4800 electron microscope was used. Observations of samples cross-sections were performed on samples freshly broken, followed by a slight platinum deposition to allow electric charges evacuation.
For contact angle measurements, a GBX-Digidrop Romans commercial apparatus was used. It allows to capture pictures of the drop at different times from a point situated in the prolongation of the sample (planar) surface. Contact angles were determined as the average between the “left” and “right” angles and together with the volume of the drop, were numerically obtained from the pictures, as a function of time. 
Photocurrent measurements were performed using obtained samples as photoelectrodes of two-electrode photoelectrochemical cells (PEC). The PEC arrangement used for photoelectrode characterization has been previously described \cite{Zumeta03}. PEC spectral photocurrent was measured in a system with a 150 W halogen-xenon lamp coupled to a monochromator LOMO MDR-12. A calibrated silicon photodiode was used for measuring incident monochromatic light intensity, which is necessary to determine both spectral response and external quantum efficiency (IPCE). This photodiode has an enhanced spectral response in the blue, violet and near UV. At constant illumination (non-modulated light intensity) the photocurrent was measured around spectra peak values using an Agilent 34410A multimeter. Only signals with higher photocurrent values could be measured for constant illumination due to the presence of noise. To measure low photocurrent values, i.e. to obtain the whole spectrum, a lock-in amplifier SR510 modulated at $14\ Hz$ was used instead of the multimeter. However, the response time of PECs was found to be larger than $0.05\ s$; therefore measurements performed at $14\ Hz$ differ from those at constant illumination. Because of this, external quantum efficiency values are reported for spectra peak values and spectral response corresponding to $14\ Hz$ modulated light intensity. In all measurements, care was taken to reproduce equal illumination conditions on the photoelectrode and on the photodiode to avoid errors due to different illumination intensity.
\section{Results and discussion}
\subsection{Morphological characterizations}
\subsubsection{Optical microscopy}
\begin{figure*}
\includegraphics[width=43mm,height=40mm]{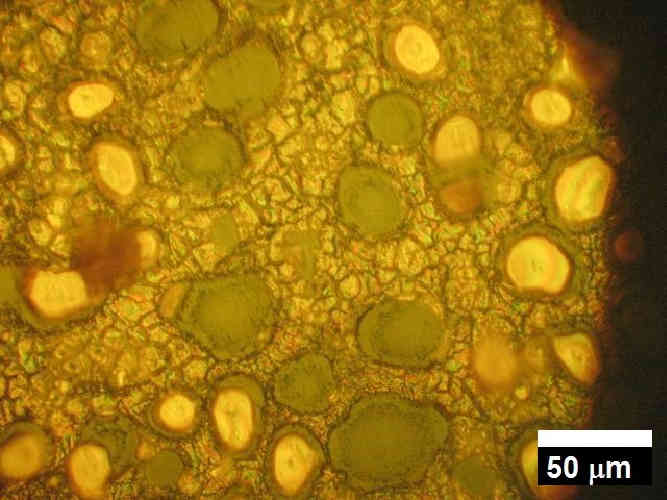}
\includegraphics[width=43mm,height=40mm]{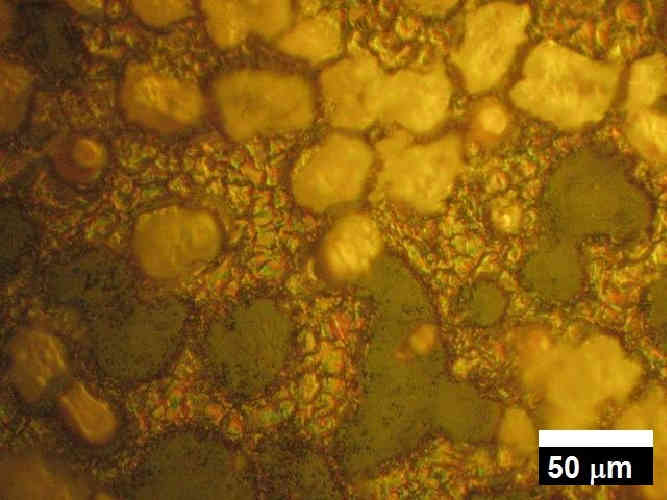}
\includegraphics[width=43mm,height=40mm]{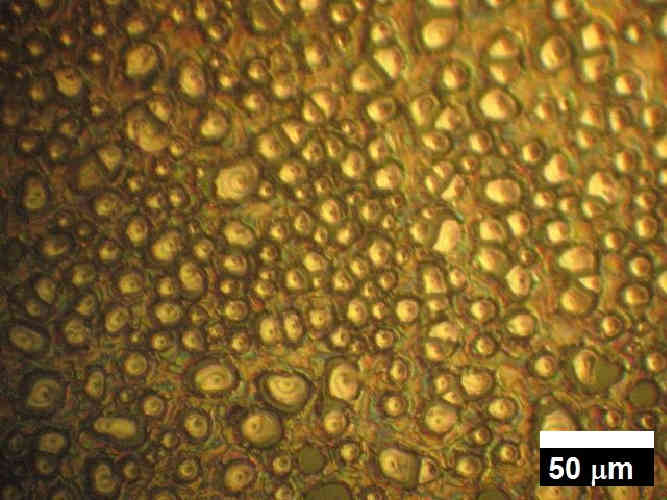}
\includegraphics[width=43mm,height=40mm]{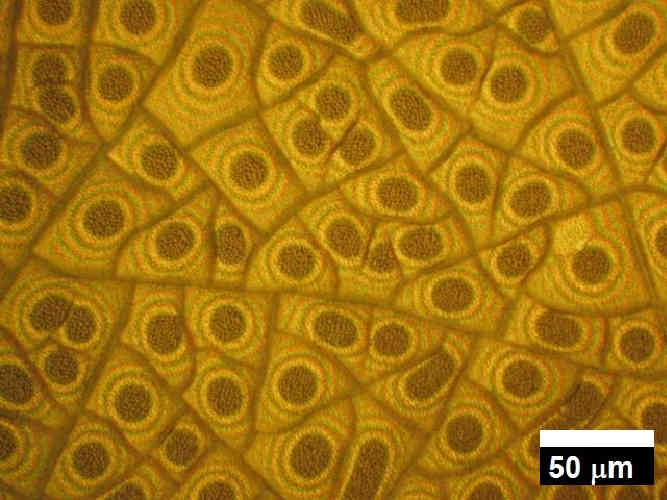}
\caption{Optical microscope images of the four studied samples corresponding, from left to right to preparations A, D, L and S.}
\label{fig:1}       
\end{figure*}
Samples A and D look rather similar, marked by the presence of agglomerates separated by regions with a different appearance. Samples L differ from the previous two by the smaller size of the agglomerates. Addition of latex seems to decrease the size of agglomerates. This is probably due to the surfactant agent in which the latex is dissolved that contributes to a better wetting and break-up of agglomerates. Surface of samples S looks very different. It is formed by compactly packed scales that cover the whole surface. In Fig. 1d) one can observe that each scale has a black elliptical center surrounded by colored rings identified as Newton rings. It seems that each scale is firmly attached in the center and Newton rings reveal that they tend to separate towards the periphery. Straight boundaries and alignment of scales are observed probably due to instability of the $TiO_2$ film-substrate interface that causes this ordered microscopic cracking. Cracking can be explained by a reduced plasticity of the thin films under drying stresses, due to the presence of a polymeric silica network linking together the anatase nanoparticles.
\subsubsection{Scanning electron microscopy}
SEM was also used in order to study materials morphology. Figure 2 shows a sequence of pictures for the four studied samples, taken at the same magnification. 
\begin{figure*}[!htb]
\includegraphics[width=42mm,height=40mm]{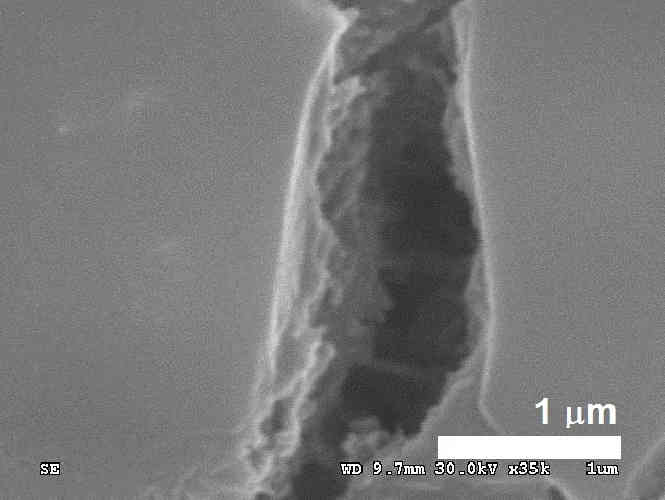}
\includegraphics[width=42mm,height=40mm]{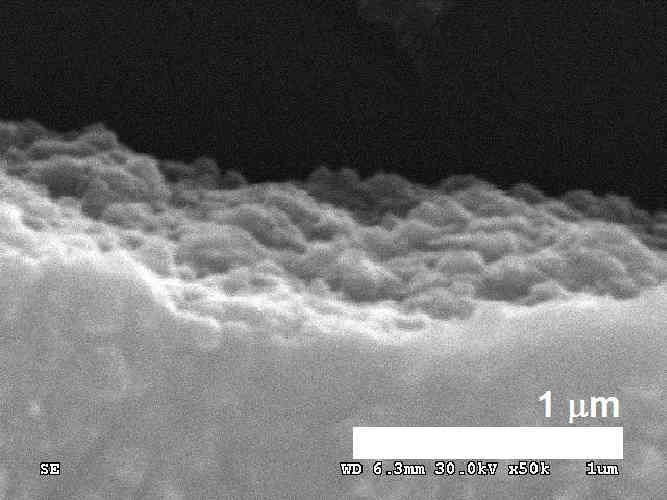}
\includegraphics[width=42mm,height=40mm]{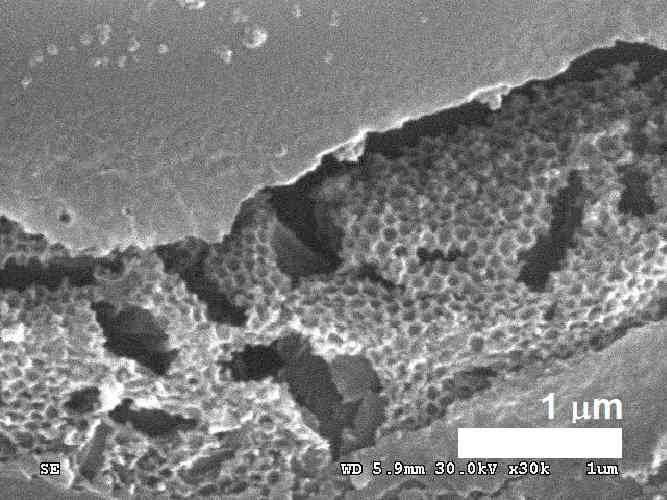}
\includegraphics[width=42mm,height=40mm]{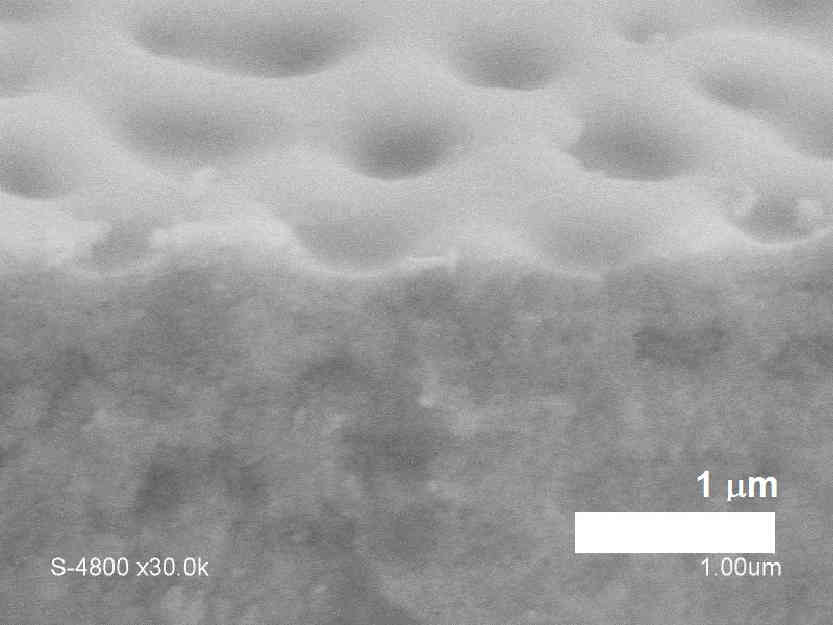}
\caption{SEM pictures of the $TiO_2$ samples obtained. From left to right, pictures correspond to samples A, D, L and S.}
\label{fig:2}       
\end{figure*}
Again samples A and D differ little when observed using SEM. Comparison of Fig. 2a) and Fig. 2b) shows that sample D is rougher than sample A. This can be explained by the incorporation of nanocrystals from  P25 (having a mean diameter of 25 nm) in sample A preparation which is known to give rise to nanocrystals of less than 10 nm diameter \cite{Bosc04}. What is even more interesting is the very particular structure observed in Fig 2c) for samples in which latex was added. It seems that when the $TiO_2$ surrounds the latex in the preparation before thermal treatment, self-organization of latex balls occurs in a rather regular way. This gives rise after calcination to a network of densely-packed pores with diameters of 50-100 nm. Sample S (Fig. 2d) clearly shows a different organization, where two types of porous matrix can be observed. One is formed by big pores of 400 - 500 nm in diameter, attributed to holes left after departure of the spherical micelles. The surrounding matter is composed of small grains, of typical size around 8 nm giving rise in the space left between them, to very small pores. The introduction of $20\%$ of $SiO_2$ in the $TiO_2$ leads to a more regular mesoscopic organization, which recalls very regular organization observed in pure $SiO_2$ \cite{Huang04}.
Film thickness was determined using cross section SEM images. Thickness values are 1.7, 1.0, 0.3 and 2.5 $\mu m$ for samples A, D, L and S, respectively. The thickness difference will be taken into account when making comparisons among samples light-response.
\subsection{Spectral response and external quantum efficiency (IPCE)}
Spectral response of all four types of samples obtained from the short-circuit spectral photocurrent is shown in Fig 3. The spectral range in which the samples respond is quite similar. Due to the low current values, this spectral response is measured modulating incident light intensity at $14$ Hz in order to eliminate noise by using a lock-in amplifier. This means that in each cycle the sample is illuminated during ca. $0.036 s$.
\begin{figure}
\resizebox{0.75\columnwidth}{!}{%
  \includegraphics{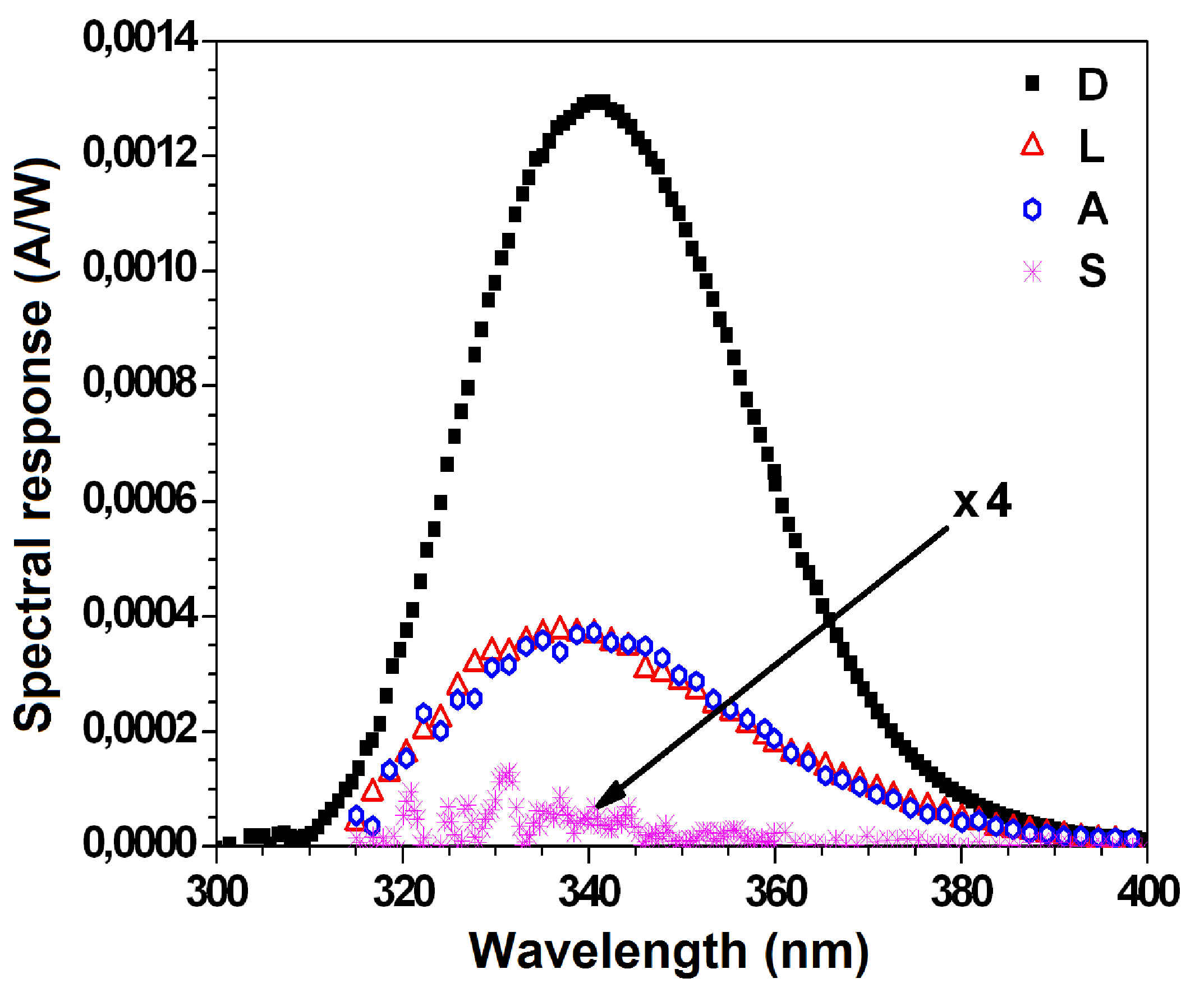}
}
\caption{Spectral response of all samples for light modulated at 14 Hz. Graphs correspond to: D - solid squares; A - open circles; L – open triangles, S – asterisks.}
\label{fig:3}       
\end{figure}
As can be observed in Fig. 3, the highest spectral response values correspond to sample D that was obtained by adding  P-25 to the sol-gel with the templating agent. A and L show similar spectral response, smaller than that of D. Sample S presents a very small spectral response, that was attributed to sample microcracks. 
Relative spectral response values could change if response times depend on sample type and they are longer than 0.036 s, i.e., relative spectral response values might be different for constant illumination. Only sample D spectral response could be measured for both, 14 Hz modulated light intensity as well as for constant light intensity in the whole spectral range, as shown in Fig 4.
\begin{figure}
\resizebox{0.75\columnwidth}{!}{%
  \includegraphics{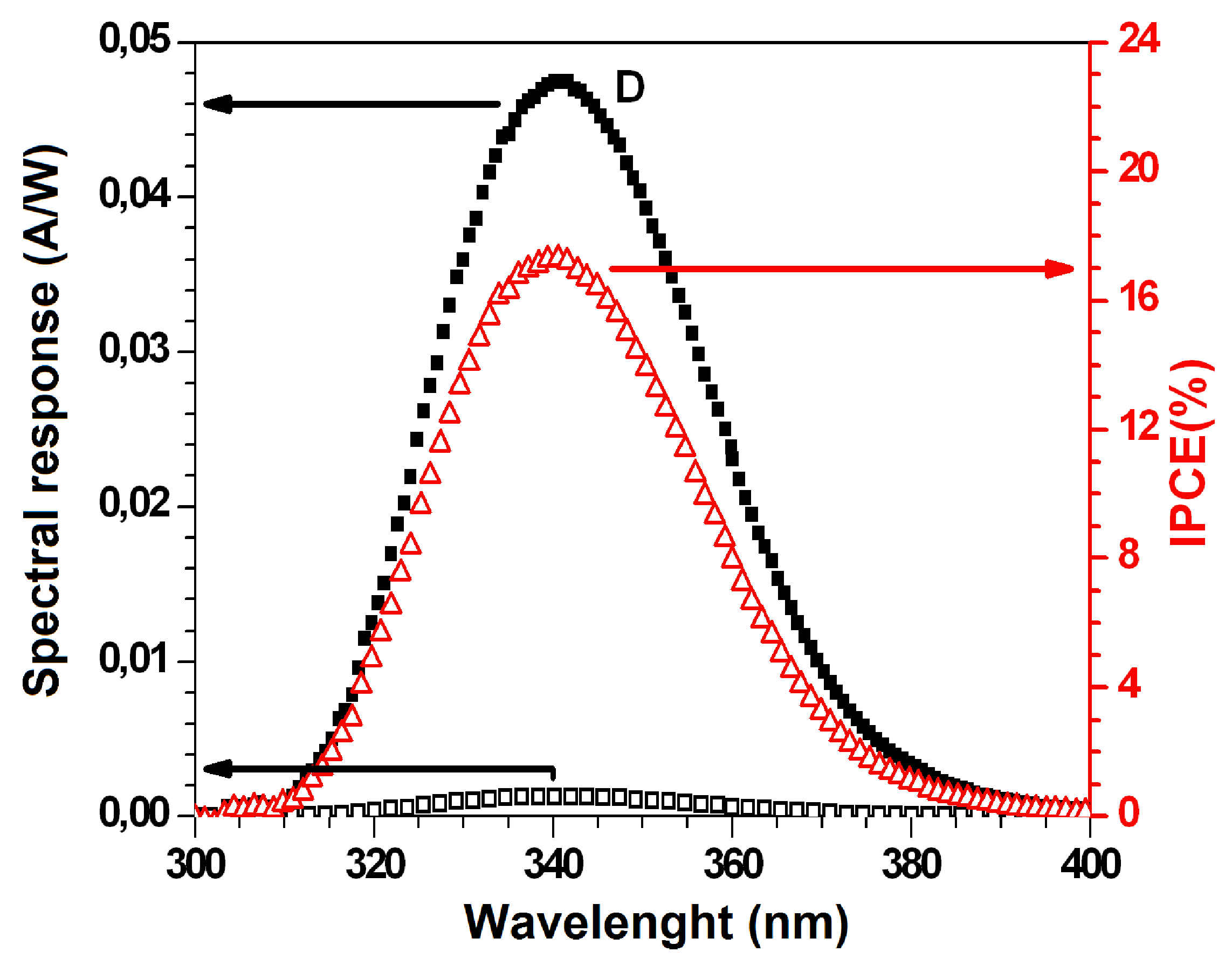}
}
\caption{Spectral response and IPCE of sample D. Filled squares refer to spectral response obtained under continuous light illumination, from which IPCE triangles were obtained. Open squares is the spectral response at 14 Hz modulated light.}
\label{fig:4}       
\end{figure}
Sample D spectral response for constant illumination is one order of magnitude larger than at 14 Hz (see Fig. 4). This confirms a response time significantly longer than 0.04 s. For other samples it was possible to compare the modulated and non-modulated maximum spectral response values. After determining peak wavelength for modulated spectral response, continuous light intensity measurement was performed at this wavelength. Samples maximum spectral response value for constant illumination is shown in Table 1. External quantum efficiency or IPCE can be obtained from the spectral response values and it is also shown in Table 1. 
In order to compare films performance, differences in sample thickness and morphology must be taken into account when considering the problem of charge transport. The mechanism for electron transport within the $TiO_2$ layer is a difficult problem that still gives rise to debates, an important question being what is the mechanism leading to an electron mobility smaller by several orders of magnitude in mesoporous titanum dioxide, than in $TiO_2$ single crystals. Besides, other particular questions are what is the exact role of sub band gap states (with a multiple-trapping charge transport involving them) and the nature of traps (located at grain boundaries or at grain surface) that limit the electron transport.
The microscopic mechanisms are out of the scope of the present paper, and we attempt here a simple phenomenological model, giving estimates of the behaviour, in order to better understand the results.
 It can be considered that the measured photocurrent coming from the sample is in first approximation proportional to:
\begin{equation}
I \sim  \omega \frac{S_{real}}{R}     
\end{equation}
where $\omega$ will be called effective interface coefficient and is meant to be the fraction of the real area that is effective in extracting holes from the $TiO_2$ to the electrolyte. That is, the $TiO_2$-electrolyte three dimensional interface ($S_{interface}$) is not equal to the real surface ($S_{real}$) because pores might be too small to allow penetration or movement of electrolyte ions. $S_{real}$ designates the $TiO_2$ surface within the layer, and R is an effective electrical resistance of the $TiO_2$ layer that depends on diffusion length and recombination time. We assume that light scattering and electron transfer from the $TiO_2$ to the FTO are second order effects. 
To estimate the samples real surface, the geometrical organization of the nanoparticles in the different samples is schematically depicted in Fig 5.
\begin{center}
\begin{figure*}
\includegraphics[width=160mm,height=45mm]{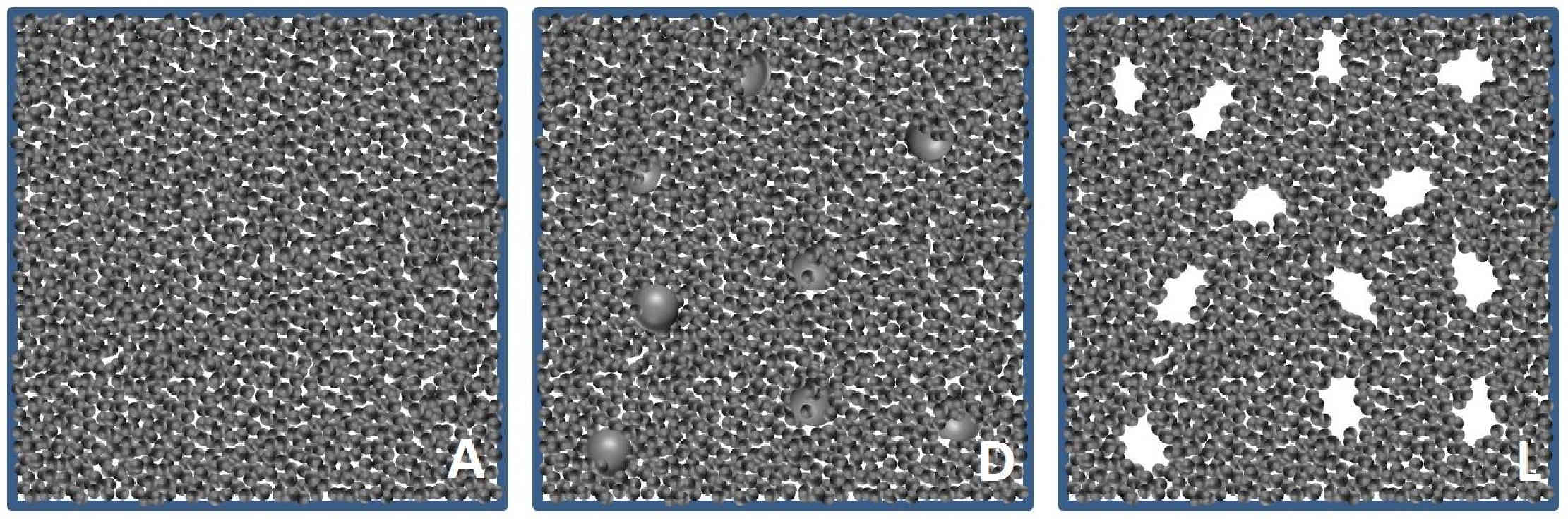}
\caption{Schematic representation of the diverse morphologies obtained in the studied preparations}
\label{fig:5}       
\end{figure*}
\end{center}
Samples A are constituted by small grains, while D samples contain $20 \%$ of  nanopowder, with average grain size of $25 nm$. Samples L are characterized by a larger porosity due to the $20\%$ of added latex particles that leave the material during the calcination process producing cavities. 
The surface $S_{real}$ of a sample of mass $m$ can be written as :
\begin{equation}
S_{real} = S_{\tiny {BET}} m
\end{equation}
where $S_{\tiny {BET}}$ is the specific surface; the mass $m$ is in turn given by:
\begin{equation}
m = (1-p) \rho_m(TiO_2) S_0 d
\end{equation}
Here, $p$ is the porosity, $\rho_m(TiO_2)$ is the mass density of pure anatase (3.84 $g/cm^3$ [ref]), $S_0$ is the visible area which is given by the light spot area (the same for all measurements) and d is the thickness of the sample. This estimation of $S_{real}$ does not take into account the contact zones between nanoparticles due to “necking” which are expected to reduce $S_{real}$. This underestimation would nevertheless introduce an error of less than $10\%$ in $S_{real}$ which would not modify the conclusion based on estimations.
For samples A we can consider $p \sim 40\%$ and $S_{BET} = 197\ m^2/g$ \cite{Bosc06}, which gives a real surface  $S_{real} [A] = 758\  cm^2$. In the case of sample D (Fig 5b), the mass is divided in very small nanocrystals with the properties of sample A (occupying $80\%$ of the total volumen) and bigger   $TiO_2$ nanocrystals dispersed in the volume of the sample and occupying $20\%$ of it. Because of this, we can consider contributions from both nanoparticles dimensions. To estimate the  nanoparticles contribution we assume that nanocrystals in  P-25 $TiO_2$ are nearly spherical with the diameter $\phi = 25\ nm$ and that they are dispersed within the D preparation without forming aggregates, which appears to be the case as observed by SEM. For spheric nanocrystals, the surface to volume ratio is equal to $6/\phi$. Considering the sum of all nanocrystals surfaces and previous assumptions, the estimated value is  $S_{real} [D] = 419\ cm^2$. Sample L has $80\%$ of small particle with the properties of sample A, so  $S_{real} [L] = 113\ cm^2$.
\begin{table}
\caption{ Samples characteristics and IPCE values}
\label{tab:1}       
\begin{tabular}{|l|l|l|l|l|}
\hline\noalign{\smallskip}
Sample  & $d(\mu m)$ & $S_{real} (cm^2$) &IPCE$(\%)$ & IPCE/$d$ \\
&&&&($d=1 \mu m$)\\
\hline
A & 1.7 & 758 & 1.4  & 0.86\\
\hline
D & 1.0 & 419 & 17.3 & 17.0\\
\hline
L &  0.3  &  113 &  1.9 & 6.1  \\  
\hline
\end{tabular}
\end{table}
In order to compare samples, values of IPCE that would correspond to 1 $\mu m$ thick samples are shown, in Table I. It is assumed that IPCE depends linearly with sample thickness. This is true when samples are thinner than optimum thickness value which is the case for our thin samples.
As can be observed, samples A have the highest  $S_{real}$ value, as expected. The introduction of $20\%$ of bigger nanoparticles into the preparation for obtaining samples D decreases the $TiO_2$ surface area. However, the response of samples D is the highest even when the IPCE peak values for $d=1\ \mu m$ are compared. Results obtained for samples L and D are interesting because with a smaller surface, they show a higher normalized IPCE in comparison with sample A. 
Considering porosity, a qualitative explanation for IPCE values can be given. Sample A has the smallest IPCE in spite of having the highest real area value. But it only has pores of $1.5\ nm$, resulting from the aggregation of the anatase nanoparticles, plus additional porosity created by the added templating agent ($<5\ nm$ pores) [6,8]. This pore size could be too small for the electrolyte species to flow easily within the mesoporous structure. This indicates that the effective area of the $TiO_2$-electrolyte interface, $S_{interface}$, is different (smaller) than the real area ($S_{real}$) of the mesoporous structure. This is confirmed by IPCE values for sample L larger than those for sample A when considering equal thickness of $1\ \mu m$. Large $100\ nm$-pores in sample L allow distribution and penetration of the electrolyte. Therefore, even though the real area of the mesoporous structure is bigger for sample A, the effective area of the $TiO_2$-electrolyte interface must be bigger for sample $L$. The larger 100 nm-pores benefit electrolyte penetration but, on the other hand, they represent quite a discontinuity that hinders electron diffusion in the $TiO_2$ toward the contact. In sample D, the higher IPCE values may originate from large $TiO_2$ nanocrystals ($\sim25\ nm$) introducing disorder in the film, creating larger pores and consequently allowing the electrolyte species to flow better within the sample. Therefore, the effective area of the $TiO_2$-electrolyte interface must be greater than in sample A. Besides, electrons do not encounter big voids that have to be surrounded like in sample L.
Taking into account that small pores hinder the penetration of electrolyte within the layer, the electrolyte-$TiO_2$ interface area, $S_{interface}$, is smaller than the real area, $S_{real}$, of the porous $TiO_2$ layer. Considering the electrical resistance in expression (1) which limits  IPCE, one can write that 
\begin{equation}
R = \rho_{el} \tau d / <S_s>
\end{equation}
where $\rho_{el}$ is the electrical resistivity, $d$ is the sample thickness, $<S_s>$ is the average cross section area of $TiO_2$ for photocarriers current, and $\tau$ depends on the tortuosity of electrons path and it may be defined as the ratio of the average real length of electron path within $TiO_2$ divided by $d$. For analyzing the problem of electron transport, since the motion takes place on the average in the direction perpendicular to the layer, we propose to consider the sample as formed by a series of nanograins rows following this direction. Then, one can deduce: 
\begin{equation}
<S_s> = ( 1-p) S_0
\end{equation}
Considering expressions (1), (4) and (5),one can write that the current ($I$) is proportional to:
\begin{equation}
I = (1-p) \omega S_{real}/\rho_{el} \tau 
\end{equation}
Therefore
\begin{equation}
\omega \sim \frac{I}{S_{real}}\ d\ \rho_{el} \frac{\tau}{1-p}
\end{equation}
assuming that $\rho_{el} \tau$ has approximately the same value for all samples since all of them involve at a microscopic level, $TiO_2$ nanograins in contact with each other, we get:
\begin{equation}
\omega \sim \frac{I}{S_{real}} d/(1-p) = \omega^{*} 
\end{equation}
We present in Table 2 values of $\omega^{*}$ for samples A, D, L:
\begin{table}
\caption{Effective interface coefficient $\omega^{*}$ according to expression (8)}
\label{tab:2}       
\begin{tabular}{|l|l|l|l|l|}
\hline\noalign{\smallskip}
Sample  & $d (\mu m)$ & IPCE/$S_{inter}$x($10^5$) &1-$p$$(\%)$ & $\omega^{*}$x($10^5$)\\
\hline
A & 1.7 & 188.7 & 60  & 525\\
\hline
D & 1.0 & 4125.6 & 68 & 6188\\
\hline
L &  0.3  &  1668 &  48 &1084  \\  
\hline
\end{tabular}
\end{table}
We see that $\omega^{*}$ has the smallest value for sample A (525 x $10^{-5}$), which means that the electrolyte would poorly interact with the $TiO_2$ surface, while it is approximately twice larger for sample L, and more than ten times larger for sample D. Samples D would hence have the $TiO_2$ mesoporous layer in which the effective interface area is the largest. 
In the above considerations, we did not take into account the effect of diffusion length nor of the recombination time, nor of light scattering but we think that the resistive mechanism approach leads to an estimation that should be relevant in first approximation. To check our qualitative explanation for the different IPCE, we have looked for information on the wetting properties and capillarity of the structures by performing the contact angle measurements described below. 
\subsection{Contact angle measurements}
Contact angle and capillarity are related phenomena. In porous materials, where liquid deposited on the surface can percolate inside the pores, the volume of the deposited drop will diminish as a function of time as a result of two mechanisms, namely liquid evaporation, and liquid penetration within the layer. Since external liquid evaporation is expected to occur with the same time constants in all cases, the differences of rate of drop volume change among samples, if any, can be attributed to the penetration mechanism. However, it is rather difficult experimentally, to estimate the volumes with accuracy by use of image processing, whereas the measurement of contact angle is easier and more accurate. Contact angle should depend on the given porosity and it must also decrease as a function of time, as a consequence of the volume variation. We could then expect a behavior similar to the one depicted in Fig. 6. \cite{Hilden03}.
\begin{figure}
\resizebox{0.75\columnwidth}{!}{%
  \includegraphics{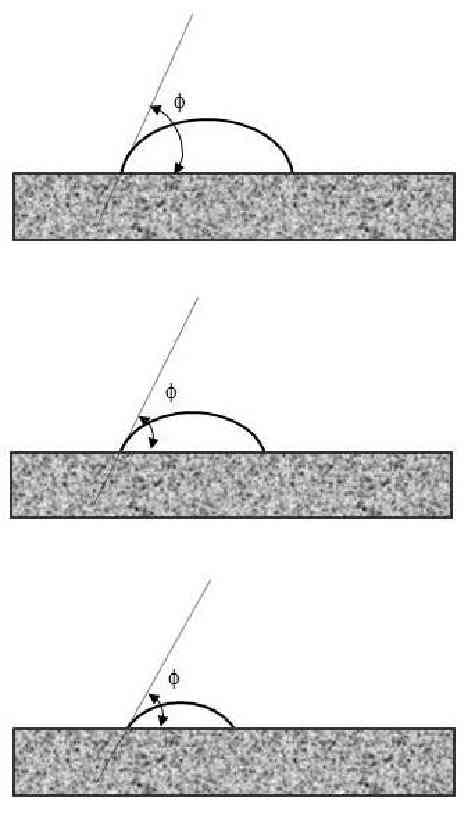}
}
\caption{Schematic evolution in time of the behavior of a drop in contact with a porous substrate showing the decrease of drop volume and contact angle.}
\label{fig:6}       
\end{figure}
Fig. 7 shows contact angle measurements for the different types of samples studied. After the initial measurement at 5 s, the contact angles diminish at a faster rate than evaporation would induce, which is attributed to the penetration of the liquid into the layer. In these conditions, a pronounced angle decreasing versus time expresses the facility for the liquid to penetrate into the layer. According to this, it is in sample D that the liquid penetrates more easily into the layer, followed by A and L which have similar results, which is in agreement with the previous presented analysis.
\begin{figure}
\resizebox{0.75\columnwidth}{!}{%
  \includegraphics{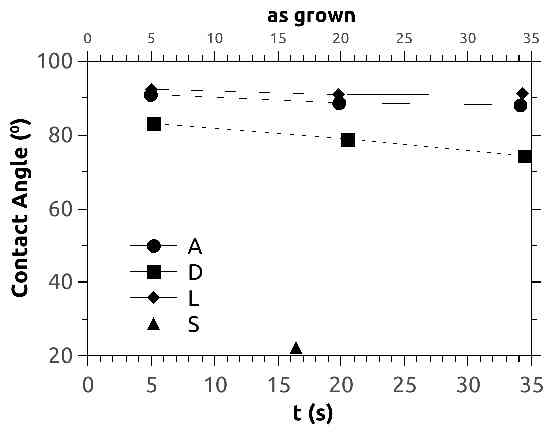}
}
\caption{Contact angle evolution with time for A, D, L and S samples. }
\label{fig:7}       
\end{figure}
\section{Conclusions}
Mesoscopic $TiO_2$ layers with different characteristics as specific area, porosity and  pore dimension have been compared. $TiO_2$ photoelectrodes on conducting glass substrates are fabricated with these layers and they are used in a two-electrode photoelectrochemical cell to measure the corresponding spectral response and IPCE. To increase efficiency of the photoelectrochemical cell, the area of the electrolyte-$TiO_2$ three-dimensional interface must be maximized, which usually implies using a $TiO_2$ film with a specific area as high as possible. However we have found that a very high specific area can hinder electrolyte penetration in the pores formed by $TiO_2$ nanograins due to very small nanocrystals and small pores. Results show that a nanocrystal network with a large specific area but that exhibits a mean pore size of 1.5 nm, resulting from the aggregation of the anatase nanoparticles, plus additional porosity created by the added templating agent ($<5$ nm pores) does not allow the electrolyte to penetrate effectively. Thus the area of the $TiO_2$-electrolyte interface is smaller than the mesoporous $TiO_2$ real area in this case. Therefore, efficiency does not necessarily increase with specific and real area increase. When larger nanocrystals or voids were added in order to create larger pores and their interconnections, the real area decreased but the IPCE values increased. These results confirm that pore size is essential in defining the actual area of the semiconductor-electrolyte interface.  
Contact angle measurements performed were in agreement with previous arguments. 
\section{Acknowledments}
We are grateful to the French Embassy in Havana for financing mobility of cuban researchers in France ("FSP Cuba", project Enermat) and to the Conseil R\'egional Languedoc Roussillon for the financement of invited professor positions in the University of N\^imes.
%
%
%

\end{document}